\documentclass[aps,prb,twocolumn,groupedaddress,footinbib]{revtex4-1}
\usepackage{bbm}
\usepackage{amsmath,leftidx}
\usepackage{graphicx}
\usepackage{times}
\usepackage{CJK}
\usepackage{color}

\begin{document}


\begin{CJK*}{UTF8}{}
\title{Kubo's response theory and bosonization with a background gauge field and irrelevant perturbations}
 \CJKfamily{bsmi}

\author{Yoshiki Fukusumi }
\author{Osor S. Bari\v si\' c}
\affiliation{}



\date{\today}

\begin{abstract}
Using conformal field theory calculations of the energy spectrum, within the XXZ model we investigate effects of the flux insertion and the Umklapp term. We discuss two approaches to the evaluation of the Drude weight, the first corresponding to the linear response theory and the second corresponding to the twisted boson theory with the Umklapp term. Divergences obtained in the context of the former contradict the Bethe ansatz results, with the two approaches coinciding for the free fermion point only. The origin of this discrepancy is in the different order in which the Umklapp term and the flux insertion are treated, where the marginal perturbation should be considered before the irrelevant. We calculate the scaling of the conductivity with system size and temperature in the long-wave limit. 
\end{abstract}

\pacs{}

\maketitle
\end{CJK*}

\section{Introduction}

A unified understanding of nonequilibrium physics is still one of the most significant and difficult tasks, involving various fundamental aspects of physical modeling. Notably, this problem can be dated back to works by Kubo, Nakano, and others\cite{doi:10.1143/JPSJ.12.570, 10.1143/PTP.15.77}. The key achievement of these pioneer investigations is the so-called (linear) response theory, which has been established around the 1950s and 1960s. From a modern perspective, Kubo's response theory can be seen as the perturbative treatment of a gauge field or a twist, being proven valid for the free fermion point \cite{2020arXiv200310390W,2020arXiv200404561W}.

On the other hand, the exact treatment of interacting one-dimensional systems with the twist or the flux by the use of the Bethe ansatz has been developed in 1980s and 1990s \cite{ALCARAZ1988280,Klumper_1991,Kitazawa_1997}. Hence, one can say that the arguments of Kubo's response theory have been established much earlier than the investigations of the specific interacting models, such as the XXZ spin chain with anisotropic parameter $|\Delta|<1$. Comprehensive analysis of this problem in Refs.~\cite{SUZUKI1971277,2003PhRvB..68m4436H, 2000LNP...544....3S} discuss conditions under which Kubo's response theory becomes correct.

In the context of nonequilibrium physics, the validity of the thermalization hypothesis is one of the most common problems for Kubo's argument. In particular, here we consider the validity of the perturbative treatment of the flux for integrable systems, keeping in mind that the integrable systems cannot reach the thermal equilibrium \cite{SUZUKI1971277}. Hence, in our case, Kubo's argument cannot be applied, at least not straightforwardly. It has been argued previously that this problem may be overcome by assuming adiabatic processes or by considering nonequilibrium steady states with open boundary conditions \cite{2000LNP...544....3S}. 

Even when one assumes the thermalization, the perturbative treatment of the flux is questionable in the context of renormalization group (RG) arguments, as discussed by van Kampen \cite{Kampen_1971}. This problem has not captured enough attention for several reasons.
As it has been emphasized previously \cite{2019PhRvB.100g5105Y}, the theories treating systems with twisted boundary condition have been developed in very different fields with weak mutual interaction, ranging from high energy and mathematical physics \cite{AlvarezGaume:1987vm,ALCARAZ1988280,Klumper_1991} to the condensed matter physics \cite{Kitazawa_1997}. Moreover, regarding calculations of conductivity, as one of the fundamental transport properties, it is difficult to follow results across the literature because various definitions are used. For example, some are based on the charge (spin) stiffness, while some are based on the Drude weight and the Kubo's response theory, whose equivalence is not easy to establish \cite{2003PhRvB..68m4436H,2000PhRvB..6111282P,2011PhRvB..83c5115S,2021RvMP...93b5003B}.

The spin stiffness for the XXZ chain \cite{PhysRev.133.A171} has been considered by Zotos {\it et al.} \cite{PhysRevLett.82.1764, 2011PhRvB..84o5125H}. Various analyses have followed, based on the Bethe ansatz and the field theory. Recently, the previous results have been reconsidered by using T- and Y- system \cite{2019arXiv190411253K}. Nevertheless, the low-temperature behaviors of the spin stiffness are still controversial. In Ref.~\onlinecite{2001EPJB...24...77L}, the finite size effects at zero temperature have been investigated by using the field theory and the Bethe ansatz. Based on the observation of the conformal invariance in the system, it has been proposed that these finite size effects may be related to the finite temperature effects for the infinite system \cite{PhysRevLett.73.332, 1998NuPhB.522..533L}. On the other hand, Fujimoto and Kawakami have proposed similar low-temperature behaviors of the Drude weight by using Kubo's
argument \cite{2003PhRvL..90s7202F}. However, even with all these efforts undertaken and regardless of their major importance, we are not aware of a work that would fully clarify the scaling behavior for both, the chain length and the temperature.

In this work, we consider the energy spectrum of the XXZ chain in the long-wave and low-temperature limit, described by the twisted boson theory with irrelevant perturbations. We compare calculations based on Kubo's argument with the results obtained by the twisted boson theory. Importantly, we find that these two approaches give different scaling behaviors. Furthermore, we provide a simple physical explanation of this discrepancy, by noticing that we are dealing with a quantum field problem with a marginal and an irrelevant perturbation. For the XXZ chain, the marginal perturbation corresponds to flux and the irrelevant perturbation corresponds to the Umklapp term. To treat this problem properly, one needs to be careful with the question of which term should be considered first.

Within the twisted boson theory, it seems natural to consider the effects of the marginal perturbation first, because the effects of irrelevant terms are treated after the considerations of the fixed point and the marginal operator. However, Kubo's argument involves the opposite ordering, considering the effect of the flux after considering the effect of irrelevant terms. Hence, except for the free fermion point that does not contain the Umklapp term, the results of the two approaches cannot match.

\section{Drude weight and spin stiffness}

We start with a short discussion of the two ways of calculating the Drude weight commonly found in the literature. In particular, we compare the Drude weight obtained from the definition of the spin stiffness and the Drude weight obtained perturbatively within the linear response theory \cite{doi:10.1143/JPSJ.12.570,10.1143/PTP.15.77,PhysRev.133.A171}. In this context, we analyze the spin chain system described by the anisotropic Heisenberg model. As well known, this problem can be mapped to the spinless fermion model using the Jordan-Winger transformation. In the presence of a magnetic flux $\Phi$, the Hamiltonian corresponding to the spin-$1/2$ anisotropic XXZ spin chain is given by,

\begin{equation}
\hat H(\Phi)=J\sum_{i=1}^{L} \left(S_{i}^{+}S_{i+1}^{-}e^{i\frac{\Phi}{L}} + h.c. +\Delta S_{i}^{z}S_{i+1}^{z}\right)\;.\label{Ham}
\end{equation}

\noindent Here, $L$ is the system length. $\Delta$ is the anisotropy parameter, directly related to the parameter of the Tomonaga-Luttinger liquid (TLL), $K$, with $\Delta=\cos(\gamma)$, $K=\pi/2(\pi -\gamma)$ and $J=(\pi-\gamma)/(2\pi\sin(\gamma))$\cite{1998NuPhB.522..533L}. For the Fermi velocity, $v_{F}\sim J$, we take the most simple normalization $v_{F}=1$. The Luttinger parameter characterizes the universal behaviors of the system such as the correlation functions and the central charge. $K=1/2$ corresponds to the $SU(2)_{1}$ Wess-Zumino-Witten model, while $K=1$ corresponds to the free Dirac fermion model \cite{Giamarchi}. Here, we concentrate on the critical phase $|\Delta|< 1$.

The Drude weight may be obtained from the spin stiffness \cite{PhysRev.133.A171,PhysRevLett.82.1764},

\begin{equation}
D_{s}=L\sum_{n} p_{k} \frac{d^{2}E_{k}^{\text{ex}}(\Phi)}{d\Phi^{2}} |_{\Phi=0},
\label{spin-stiffness}
\end{equation}

\noindent where $E^{\text{ex}}_{k}(\Phi)$ are the eigenenergies of the Hamiltonian with flux $\Phi$ and $p_{n}$ are the Boltzmann weights. The solutions of the problem with finite flux may be obtained from the Bethe ansatz or by using the field theory approach \cite{ALCARAZ1988280,Klumper_1991,Kitazawa_1997}.

By treating the flux $\Phi$ as a perturbation, the Hamiltonian (\ref{Ham}) is written in the form of the perturbative expansion \cite{PhysRev.133.A171,2020arXiv200310390W},

\begin{equation}
\hat H(\Phi)=\sum_{n=0}\frac{1}{n!}\left(i\frac{\Phi}{L}\right)^{n}\hat H_{n},
\label{flux}
\end{equation}

\noindent where $\hat H_{0}=\hat H(0)$, while for $n\ge 1$ one has $\hat H_{2n-1}=-i\hat j=J\sum_{i=1}^{L}S^{+}_{i}S^{-}_{i+1}-S^{-}_{i}S^{+}_{i+1}$, $\hat H_{2n}=\hat T=J\sum_{i=1}^{L}S^{+}_{i}S^{-}_{i+1}+S^{-}_{i}S^{+}_{i+1}$. (For the readers familiar with conformal field theory (CFT), it should be noted that we use $\hat{T}$ as the kinetic energy of the model not as the energy-momentum tensor.) Thus, up to the second order in this perturbation theory, the energy of the system is given by,

\begin{eqnarray}
E_{k}^{\text{Kubo}}(\Phi)&=&\langle k|\left(\hat{H}(0)+\frac{\Phi}{L}\hat j-\frac{1}{2}\left(\frac{\Phi}{L}\right)^{2}\hat T\right)|k\rangle\nonumber\\
&-&\left(\frac{\Phi}{L}\right)^{2}\sum_{k'\neq k}\frac{|\langle k' | \hat j|k\rangle|^{2}}{E_{k}^{\text{ex}}(0)-E_{k'}^{\text{ex}}(0)}\;,
\label{naive}
\end{eqnarray}

\noindent where $k$ labels the eigenenergies and eigenstates of the Hamiltonian without flux $\Phi=0$ again. Many works are based on this perturbative approach \cite{PhysRev.133.A171,2011PhRvB..84o5125H,2019arXiv190411253K}, since it permits to analyze the systems for finite flux $\Phi$ by calculating the matrix elements and the energies for the system without flux. 

Inserting the energies obtained perturbatively, instead of the exact solutions in Eq.~(\ref{spin-stiffness})  one obtains another expression for the Drude weight,

\begin{equation}
D=\frac{1}{L}\left( \langle-\hat{T}\rangle_{\text{th}}-2\sum_{k\neq k'}p_{k'}\frac{|\langle k' | \hat{j}|k\rangle|^{2}}{E_{k}^{\text{ex}}(0)-E_{k'}^{\text{ex}}(0)}\right)
\end{equation}

\noindent where $\langle .\rangle_{\text{th}}$ denotes the thermal average.

Frequently \cite{PhysRev.133.A171,PhysRevLett.82.1764,2020arXiv200310390W}, it is assumed that the two expressions (\ref{flux}) and (\ref{naive}) give the same result, $D=D_{s}$. However, this statement is not self-evident, particularly for low temperatures, when the system is described by TLL. 

Few problems should be emphasized. First, Eq.~\eqref{naive} is only valid for free fermions. It is inconsistent with the Bethe ansatz and the field theory with interaction and finite flux. The free fermion point is special because Eq.~\eqref{naive} becomes exact if the perturbative theory is taken into infinite order in $\Phi$. 

For free fermions, each point is two-fold degenerate by momentum inversion. This degeneracy is slightly (or "irrelevantly") broken by the Umklapp term because this term prevents the momentum from being a good quantum number. In other words, the $U(1)$ current $\hat j$ cannot become conserved, except for the free fermion point. Moreover, because the two-fold degeneracy is broken, the second term in Eq.~\eqref{naive} may grow as the limit $L\rightarrow \infty$ is approached.

\section{Calculation of spin stiffness and Kubo-Drude weight without Umklapp term}

According to the numerical and analytical findings \cite{CARDY1986186,Kitazawa_1997}, the spectrum of the system in the long-wave and low-temperature limit is given by,

\begin{equation}
E_k(\Phi)-E_0=\frac{2\pi}{L}\left(K\left(m+\frac{\Phi}{2\pi}\right)^{2}+\frac{n^{2}}{4K}+N+\overline{N} \right)\;,
\label{fluxenergy}
\end{equation}

\noindent with the ground-state energy

\begin{equation}
E_0=-L\frac{2(\pi-\gamma)}{\pi^{2}}\int^{\infty}_{0}\frac{\text{sin}\frac{\gamma t}{\pi} }{\sinh(t)\cosh(\frac{\pi -\gamma}{\pi}t)}-\frac{\pi}{6L}\;.
\label{gs}
\end{equation}

\noindent $E_k(\Phi)$ are the energy of the excited states. The index $k$ denotes different combinations of four indicies characterizing the type of exciations, $k\equiv\{n,m,N,\overline{N}\}$. $m$ denotes the integer momentum, $n$ is the magnetization index \cite{1998NuPhB.522..533L, PhysRev.150.327}, while $N$ and $\overline{N}$ are positive integers, specifying the particle-hole excitations with degeneracy $g(N)$ and $g(\overline{N})$ \cite{2012PhyE...46..119M, Klumper_1993}. Hereafter, we consider the system with zero magnetization fixing $n=0$, or the half-filled case in the spinless fermion representation of the model. It should be noted that the two-fold degeneracy, $m\rightarrow -m$, is removed by the flux, even in the absence of the Umklapp term.

In the context of the CFT, each primarly state of the system is represented by,

\begin{equation}
|m,n\rangle_{\Phi} =\exp{\left(i\left(m+\frac{\Phi}{2\pi}\right)\phi(x) +i n \theta (x)\right)}|0\rangle\;,
\end{equation}

\noindent where $|0\rangle$ is the vacuum state, while $\phi(x)$ and $\theta(x)$ are the dual bosonic fields, satisfying the commutation relation, $\left[ \phi (x), \partial \theta (x') \right]=i\delta(x-x')$. By applying descendant field (or particle-hole excitation operators) on the vacuum state, one obtains descendant states $|m,n,N,\overline{N}\rangle$ with the energy given by Eq.~\eqref{fluxenergy}. These states are eigenstates of the total momentum $\hat j$, 

\begin{equation}
\hat j|m, n\rangle_{\Phi} =2K\left(m+\frac{\Phi}{2\pi}\right)|m, n\rangle_{\Phi}\;.
\end{equation}

\noindent Consequently, the last term in \eqref{naive}, involving non-diagonal processes, gives a finite contribution only if the presence of the Umklapp in the model. 

The interesting question that remains unanswered is whether or not in the absence of Umklapp the two expressions for the Drude weight (\ref{flux}) and (\ref{naive}) coincide. We argue, using the CFT, that this is not the case. In particular, from Eq.~(\ref{fluxenergy}) one may easily calculate $D_{s}$, 

\begin{equation}
D_{s}=L\sum_{k} p_{k} \frac{d^{2}E_{k}}{d\Phi^{2}}=\frac{K}{\pi}\;,
\label{stiffness_free}
\end{equation}

\noindent where the normalization $\sum p_k=1$ is assumed, with the $k$-summation running over $m$, $N$, and $\overline{N}$ and with degeneracies $g(N)$ and $g(\overline N)$ accounted for. Hence, if we think about our system in terms of the TLL, which is an accurate description of the long-wave and low-temperature behavior, one obtains a temperature and $L$-independent $D_s$. In other words, for the temperature-dependent behavior, the effects of irrelevant perturbations should be taken into account. Moreover, as we will show in the next section, this finite contribution to the Drude weight at $T=0$ temperature is quite robust. This is as well consistent with the temperature dependence $D\approx D(T=0)+aT^{b}$ obtained by Zotos \cite{PhysRevLett.82.1764}.

Using the CFT further, we can calculate $D$ in Eq.~(\ref{naive}), where in the absence of Umklapp processes only the first term contributes. By considering the $U(1)$ symmetry generated by the current operator $\hat j$ and the total magnetization operator $\sum_{i}S^{z}_{i}$, for the half-filling $n=0$ case \cite{1998NuPhB.522..533L} the kinetic energy is given by $\hat{T}\sim \int dx \left(\partial_{x} \theta\right)^{2}$, 

\begin{eqnarray}
D&=&\frac{1}{L}\sum_{k}g(N)g(\overline{N})\langle -T_{m,N,\overline{N}} \rangle
=-\frac{aE_{0}}{L}-bU_{b}\nonumber\\
U_{\text{b}}&=&\frac{1}{L}\sum_{k}g(N)g(\overline{N}) \left(E_{k}-E_{0}\right)p_{k}\;,
\end{eqnarray}

\noindent where $a$ and $b$ are constants of proportionality that depend on the interaction, while $U_{\text{b}}$ is the CFT energy density of the bosonic excitations,

\begin{equation} 
U_{\text{b}}=-\frac{1}{L}\frac{\partial}{\partial \beta}\text{log}[Z_b]\;.
\end{equation}

\noindent The partition function is within the CFT given by,

\begin{equation} 
Z_{\text{b}}(\tau)=\frac{1}{|\eta(\tau)|^{2}}\sum_{m}q^{Km^{2}}\;.
\end{equation}

\noindent with $\tau=i\beta /L$, $q=e^{2\pi i\tau}$, and $\eta(\tau)=q^{1/24}\prod_{N'=1}^{\infty}(1-q^{N'})$ and $\beta$ is inverse temperature of the system \cite{Ginsparg:1988ui,2019PhRvB.100g5105Y}. In this notation, we can express the Kubo-Drude weight by, 

\begin{eqnarray}
D&\sim& - \frac{aE_{0}}{L} +\frac{ib}{L^{2}}\frac{\partial}{\partial \tau}
\text{log}[Z_{\text{b}} (\tau)]\nonumber\\
&=&-\frac{aE_{0}}{L}+\frac{ib}{\beta L}\frac{\partial}{\partial \tau'}
\text{log}[Z_{\text{b}} (\tau])\;,
\label{Drude_free}
\end{eqnarray}

\noindent with $\tau'=-1/\tau$ and where we have use the modular invariance of the model, $Z_{\text{b}}(\tau)=Z_{\text{b}} (\tau')$.

One can easily see now that the second term in Eq.~\eqref{Drude_free} gives a different scaling with $L$ in comparison to Eq.~\eqref{stiffness_free}. This difference cannot be solved by considering the effect of irrelevant perturbations, except for the free fermion point. Furthermore, we will show that the effects of the Umklapp term cannot be treated successfully by the Kubo-Drude formalism irrespectively of the interaction.

\subsection{Free fermion point}

For the free fermion point, it is easy to obtain the exact form of energy spectrum by using the Fourier transformation,

\begin{equation}
E_{k}^{\text{ex}}(\Phi)\propto - \sum_{j} \cos\left(\frac{\pi(2j+1)}{L}+\frac{\Phi}{L}\right)\;.
\label{free-fermion}
\end{equation}

\noindent By taking the derivative of Eq.~\eqref{free-fermion}, one obtains

\begin{equation}
\frac{d^{2}E_{k}^{\text{ex}}}{d\Phi^{2}}=-\frac{1}{L^{2}}E_k^{\text{ex}}(\Phi)\;.
\end{equation}

\noindent Thus, in the both cases, Eqs.~\eqref{spin-stiffness} and \eqref{naive}, in the absence of interaction the same result is obtained for any temperature, $D_s=D$. It should be noted that \eqref{free-fermion} can only contain the corrections of irrelevant terms which do not break the two-fold degeneracy of the system at $\Phi=0$. Hence there exists no contribution from the Umklapp term in this case\cite{2018PhRvB..97p5133K}.

\section{Umklapp term and Kubo's response theory}

We start our discussion of the Umklapp term by considering the zero-flux case. Since the other excitations are unimportant for the present analysis, we consider explicitly just the excitations corresponding to the momentum $m$. Without the flux, the spectrum of the system is doubly degenerate except for the ground $m=0$ state of even ($+$) parity. In the basis of odd and even states, $m>0$,

\begin{equation}
|m,\pm\rangle=\frac{1}{\sqrt{2}}\left(|m\rangle \pm |-m\rangle\right)\;.\label{basis}
\end{equation}

\noindent the Umklapp term,

\begin{equation}
V_U=\frac{\lambda}{2\pi}\int dx\;\cos[2\phi(x)]\,,\label{Uterm}
\end{equation}

\noindent introduces transitions between states of the same parity. For $m\geq2$, one gets \cite{CARDY1986186}

\begin{equation}
\langle m', \pm |\cos2\phi|m, \pm\rangle\propto\frac{\delta_{m',m+2}+\delta_{m',m-2}}{2}\left(\frac{2\pi}{L}\right)^{4K}\;, \label{MEL}
\end{equation}

\noindent while for $m=1$ one obtains,

\begin{equation}
\langle 1, \pm |\cos2\phi|1, \pm\rangle \propto\pm\frac{1}{2}\left(\frac{2\pi}{L}\right)^{4K}\;. \label{MEL1}
\end{equation}

\noindent Using the perturbation theory it is easy to calculate the change of the energy spectrum. In particular, for the first $m=1$ excited state the leading correction to the energy depends on the parity,

\begin{equation}
E^{\text{ex}}_{1,\pm}(0)-E_{1}(0)\sim\Delta_{1,\pm}=\pm\left(\frac{2\pi}{L}\right)^{4K-1}\frac{\lambda}{2}\;,\label{Delta1}
\end{equation}

\noindent and for $m=1$ the two-fold degeneracy is lifted in the first order in $\lambda$.

Since the current operator is odd under space inversion, the nonzero matrix elements involve states of opposite parity, $\langle m,\pm |\hat{j} |m' ,\mp \rangle=m\;\delta_{m,m'}$. Combining the perturbative treatment of the Umklapp term in Eq.~\eqref{Delta1} and the perturbative treatment of the flux in Eq.~\eqref{flux}, one may easily determine the contribution to the Drude weight given by the last (nondiagonal) term in Eq.~\eqref{naive}. In the low temperature limit this contribution is dominated by the lowest excitation, scaling as, 

\begin{equation}
\frac{1}{L^{2}}\frac{|\langle 1,+|\hat{j}|1,-\rangle|^{2}}{2\Delta_{1,+}}\approx \frac{1}{4\pi^{2}\lambda}\left(\frac{L}{2\pi}\right)^{4K-3}\;. 
\end{equation}

Within the perturbative treatment, the energy of the system with the flux and the Umklapp term may be expressed by,

\begin{equation}
E_{1,\pm}^{\text{Kubo}}(\Phi)\sim E_{1,\pm}(0)\mp\frac{K^{2}\Phi^{2}}{\pi^{2}\lambda}\left(\frac{L}{2\pi}\right)^{4K-3}\;.\label{eqEpm}
\end{equation}

\noindent Thus, for $K\ge 1/2$, as a function of $\Phi$ we should observe an additional energy splitting if Kubo's formulation is valid. However, such a splitting contradicts the numerical results obtained by the Bethe ansatz and the corresponding field theory \cite{ALCARAZ1988280,Klumper_1993,Kitazawa_1997}. Moreover, whereas Eq.~\eqref{fluxenergy} and the results of the Bethe ansatz contain the linear term proportional to $\Phi/L$, it is impossible to obtain such a contribution within the perturbative approach in Eq.~\eqref{eqEpm}. 

Regarding higher excitations, it follows from Eqs.~\eqref{MEL} and \eqref{MEL1} that the two-fold degeneracy is lifted by the Umklapp term for all $m$. Within the perturbation theory, for odd $m$ each state $|m,\pm\rangle$ in Eq.~\eqref{basis} is connected to the state $|1,\pm\rangle$. Unlike the odd parity states,  the even parity states are connected to the vacuum $|0\rangle$ state. In any case, the two-fold degeneracy becomes lifted in the $m$-th order in $\lambda$. The energy splitting given by the perturbative theory may be expressed by

\begin{equation}
\Delta_{m,+}-\Delta_{m,-}= c_m\lambda^m\left(\frac{2\pi}{L}\right)^{m(4K-2)+1}.
\end{equation}

\noindent where $c_m$ is the coefficient given by the $m$-th order perturbation theory in $\lambda$. However, for large $m$, the energy splitting induced by the flux $\Phi$, 

\begin{equation}
E_{m,\pm}^{\text{Kubo}}(\Phi)=E_{m,\pm}(0)\mp\frac{m^{2}K^{2}\Phi^{2}}{\pi^{2}c_m\lambda^{m}}\left(\frac{L}{2\pi}\right)^{m(4K-2)-1},
\label{Kubo-energy}
\end{equation}

\noindent grows with $L$. This means that the last term in Eq.~\eqref{naive} grows in the $L\rightarrow \infty$ limit as well, which means that the linear response theory in Eqs.~\eqref{flux} and \eqref{naive} contains a problem for finite fluxes.

\section{Twisted boson and conformal perturbation theory}

To take into account the effects of both, the Umklapp term in Eq.~\eqref{Uterm} and the flux $\Phi$ properly, we consider twisted boson theory. We use the Hamiltonian formalism of the CFT with the perturbation theory introduced by Cardy \cite{CARDY1986186}. As we have shown, the quantum states without flux and with the Umklapp term are specified by the positive index $m$ and their parity in Eq.~\eqref{basis}. On the other hand, within the twisted boson theory, the quantum states are specified by the momentum $m+\Phi/2\pi$. Hence these two formalisms cannot be connected continuously by taking the $\Phi\rightarrow0$ limit. In fact, by considering higher-order contributions in $\lambda$, one generally gets higher-order singularities in state energies.

The matrix elements in the $|m\rangle$ basis associated with the Umklapp term are given by

\begin{equation}
\begin{split}
\langle m|2V_U/\lambda|m'\rangle 
=\delta_{m+2,m'}\left( \frac{2\pi}{L} \right)^{h_{-}}+\delta_{m-2,m'} \left( \frac{2\pi}{L} \right)^{h_{+}},
\end{split}
\label{matrix}
\end{equation}

\noindent where the exponents $h_{\pm}$ are flux dependent, $h_{\pm}=4K(1\pm\Phi/2\pi)-1$, coming from the conformal dimension of $e^{\pm2\phi}$. We have dropped indices $\{ n,N,\overline{N}\}$ as in the previous section.

It follows from Eq.~\eqref{matrix} that within the perturbative theory there are no linear contributions in $\lambda$, as expected from the charge neutrality condition of the free boson CFT. In the second order in $\lambda$, the energy of the states $E_k(\Phi)$ in Eq.~\eqref{fluxenergy} gets an additional contribution $E_U$, ,

\begin{equation}
E_U=-\frac{\lambda^2}{16K}
\left( \frac{1}{m+1+\frac{\Phi}{2\pi}} -\frac{1}{m-1+\frac{\Phi}{2\pi}} \right)\left( \frac{2\pi}{L}\right)^{8K-3}\;,
\label{perturbed_energy}
\end{equation}

\noindent with $E^{\text{ex}}_{k}(\Phi)-E_{0}\sim E_k(\Phi,\lambda)=E_k(\Phi)+E_U$. Because $K\geq1/2$, in comparison to $E_k(\Phi)$ in Eq.~\eqref{fluxenergy}, $E_U$ behaves as a subleading contribution to the total energy $E_k(\Phi,\lambda)$. For $\Phi =0$, in its perturbative form in Eq.~\eqref{perturbed_energy}, $E_U$ involves a singularity for $m=\pm 1$ because we do not consider the degenerate perturbation theory. However, by summing the $m$ and $-m$ contributions in Eq.~\eqref{perturbed_energy} together, this divergent behaviour is easily removed in the $\Phi\rightarrow0$ limit.
(By considering higher-order perturbation theories, we can obtain the divergence for general $m$, but this is out of the scope of the present work.)

With the flux dependence of the energry in Eq.~\eqref{perturbed_energy} known, we can discuss in more details its adiabatic physics. The traditional discussion can be found in Appendix \ref{Kubo_appendix} or Refs.~\onlinecite{2020arXiv200310390W, 2020arXiv200404561W}.
The higher order derivatives of Eq.~\eqref{perturbed_energy}, which are closely related to higher response functions and the nonliear Drude weight investigated in Refs.~\cite{2020arXiv200310390W,2020arXiv200404561W,2021arXiv210305838T}, are given for $l>2$ by

\begin{equation}
\frac{d^{l}E_{U}}{d\Phi^{l}} =
- \left( -2\pi\right)^{-l} \frac{\lambda^{2}}{16K}(l!) A_{l}(m,\Phi )\left( \frac{2\pi}{L}\right)^{8K-3}\;,
\label{derivative}
\end{equation}

\noindent with

\begin{equation}
A_{l}(m,\Phi )=\frac{1}{\left(m+1+\frac{\Phi}{2\pi}\right)^{l+1}} -\frac{1}{\left(m-1+\frac{\Phi}{2\pi}\right)^{l+1}}\;.\label{Am}
\end{equation}

\noindent Hence, the derivatives in Eq.~\eqref{derivative} vanishe in the $L\rightarrow \infty$ limit. Actually, this behavior is consistent with the "divergence" at zero temperature of the non-linear Drude weight discussed in Refs.~\onlinecite{2020arXiv200310390W, 2020arXiv200404561W,2021arXiv210305838T}. This divergency follows from the different defintion of the non-linear Drude weight in these works, which involve the additional $L^{l-1}$ factor, $L^{l-1}d^{l}E_{0}^{\text{ex}}(\Phi)/d\Phi^{l}\sim L^{l-4-8K}$. On the other hand, other irrelevant terms which may be expressed as multiplication of holomorphic and antihormorphic energy-momentum tensors \cite{1998NuPhB.522..533L,2012IJMPB..2644009S} can only give finite contributions to the non-linear Drude weight.

In the $\Phi\rightarrow 0$ limit, the thermal average of the derivatives in Eq.~\eqref{derivative} are approximatelly given by,

\begin{equation}
\left<\frac{d^{l}E_{U}}{d\Phi^{l}} \right>_{\text{th}}=\sum_k\frac{e^{-\beta E_k(\Phi=0,\lambda=0)}}{Z_{\text{b}}(\tau)}
\frac{d^{l}E_k(\Phi,\lambda)}{d\Phi^{l}},\label{thav}
\end{equation}

\noindent where $E_k(\Phi=0,\lambda=0)=E_{k}(0)-E_{0}$ in Eq.~\eqref{fluxenergy}. The terms in Eq.~\eqref{thav} vanish for odd $l$, since in these cases the terms for $m$ and $-m$ given by Eq.~\eqref{Am} cancel. For even $l>2$, in the $\Phi\rightarrow0$ limit one gets

\begin{equation}
\begin{split}
&\left<\frac{d^{l}E_{U}}{d\Phi^{l}} \right>_{\text{th}} \\
&=-\sum_{n,N,\overline{N}}g(N)g(\overline{N}) \frac{\lambda^{2}}{16K}\left( -2\pi\right)^{-l} (l!)\left( \frac{2\pi}{L}\right)^{8K -3} \\
&\times\left( 2p_{0,n,N,\overline{N}}+\frac{p_{1,n,N,\overline{N}}}{2^{l}}+2\sum_{m>1}A_{l}(m,0) p_{m,n,N,\overline{N}}\right).
\label{result}
\end{split}
\end{equation} 

\noindent The last term in Eq.~\eqref{result},

\begin{equation}
F_{m>1}^{l}=2\sum_{m>1} A_{l}(m,0)e^{-\frac{2\pi\beta K}{L}m^{2}}\;,
\end{equation}

\noindent may be evaluated numerically. We can now express Eq.~\eqref{result} as,

\begin{eqnarray}
\left<\frac{d^{l}E_{U}}{d\Phi^{l}} \right>_{\text{th}}&=&-\frac{(l!)\left(-2\pi\right)^{-l}\lambda^{2}}{16K\;z(L,\beta)}\nonumber \\
&\times&\left(2+\frac{e^{-\frac{2\pi\beta K}{L}}}{2^{l}} + F_{m>1}^{l}\right)\left( \frac{2\pi}{L} \right)^{8K-3}
\label{scaling}
\end{eqnarray}

\noindent with the partition function $z(L,\beta)=\sum_{m=-\infty}^{\infty}e^{-\frac{2\pi\beta K}{L}m^{2}}$ taking the form of the Rieman theta function.

We can use Eq.~\eqref{fluxenergy} and \eqref{scaling} for $l=2$ to obtain the spin stiffness,

\begin{equation}
D_{s}=\frac{K}{\pi}-\frac{\lambda^{2}}{32\pi^{2}K}G(x)\left( \frac{2\pi}{L} \right)^{8K-4},
\label{CFT_Drude}
\end{equation}

\noindent where, for a given value of $K$, $G(x)$ is a function of a single parameter $x=\beta/L=-i\tau$,

\begin{equation}
G(x)=\frac{1}{z(L,\beta)}\left(2+\frac{e^{-\frac{2\pi\beta K}{L}}}{4} + F_{m>1}^{l=2}\right)\;.\label{Gx}
\end{equation}

\begin{figure}[tb]
\includegraphics[width=1.\columnwidth]{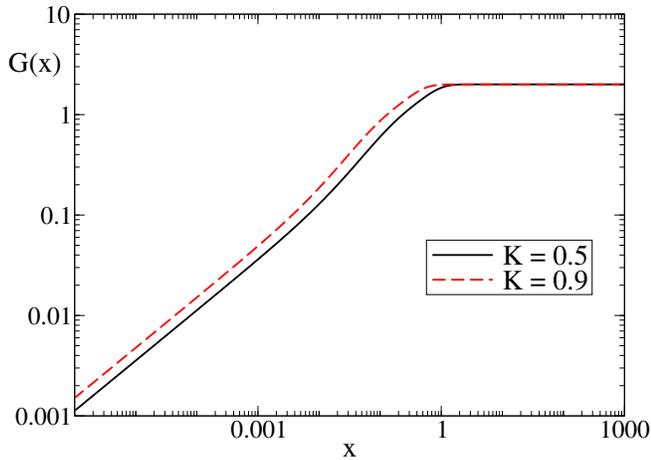}
\caption{Results showing the scaling behavior of $G(x)$ in Eq.~\eqref{Gx} for $K = 0.5$ and $K=0.9$. For small $x<0.005$, one clearly observes the power-law behavior, with the coefficient $\alpha=0.5$. This power-law behavior is governed by the $\beta/L$ dependency of the partition function $z(L,\beta)$.}
\label{fig1}
\end{figure}

The function $G(x)$ is shown in Fig.~\ref{fig1}. In the zero-temperature limit, when $x$ is large, the finite-size effects are described by a temperature-independent power-law behavior,

\begin{equation}
D_{s}=\frac{K}{\pi}-\frac{\lambda^{2}}{32\pi^{2}K}\left( \frac{2\pi}{L} \right)^{8K-4}\; ,
\end{equation}

\noindent and this result is consistent with the RG result in Ref.~\onlinecite{2001EPJB...24...77L}. Further exponentially small corrections in the low-temperature region can be evaluated from \eqref{CFT_Drude}.

In the $x\rightarrow 0$ limit, the Drude weight preserves a temperature dependence,

\begin{equation}
D_{s}=\frac{K}{\pi}-\frac{\lambda^{2}}{32\pi^{2}K}\sqrt{\frac{2K\beta}{L}}\left( \frac{2\pi}{L} \right)^{8K-4}\;.\label{ftemp}
\end{equation}

\noindent At elevated temperatures in Eq.~\eqref{ftemp}, the finite-size effects become strongly suppressed as $L$ increases. In particular, irrespectively of the value of $K$, for $x=\beta/L<0.005$ the function $G(x)$ in Fig.~\ref{fig1} follows the power-law behaviour $x^{-\alpha}$, with $\alpha=1/2$. This limit is beyond the present exact diagonalization approaches since it involves very large system sizes $L>200$ even for quite large temperatures, $\beta=1$. However, it is not clear whether the CFT can approximate the energies of the lattice models in this region. It should also be noted that the Lagrangian formalism may be more useful to analyze this limit \cite{1998NuPhB.522..533L}, especially for the calculations of the free energy and correlation functions. However, there exist ambiguities to define susceptibility in this formalism \cite{2000PhRvB..6111282P}, and the comparison between our results and those from Lagrangian formalism is a future problem.

In the intermediate region $0.005<x<1$, $G(x)$ cannot be described by the power-law function, changing quite sharply near $x\approx1$. In this intermediate region, effects of other irrelevant terms which we have neglected, for example, that of $\text{cos}4\phi$, may appear if $L$ are not sufficiently large. This may explain controversial findings of the literature \cite{2011PhRvB..84o5125H, 2003PhRvB..68m4436H,PhysRevLett.82.1764}.

As we show in the appendix \ref{TT_appendix}, there exist other irrelevant terms with conformal dimension $4$\cite{1998NuPhB.522..533L, 2012IJMPB..2644009S}.
Here we show only the results of the contributions of these terms to spin stiffness,
\begin{equation}
D_{s}^{TT}=F_{TT}+Q_{TT}+\frac{\pi K}{6L^{2}}\left(\lambda_{+}-2\lambda_{-}\right)
\end{equation}
where we have used the functions,

\begin{align}
F_{TT}&=\frac{4\pi b}{L^{2}}\sum_{m}m^{2} \frac{e^{-\frac{2\pi \beta K}{L}m^{2}}}{z(L,\beta)}, \\
Q_{TT}&=-\frac{2 b'}{L}\frac{\partial}{\partial \beta}\text{log}\frac{1}{\eta(\tau)^{2}},
\end{align}

\noindent and $b$ and $b'$ are two parameters defined by coupling constants $\lambda_{\pm}$ of these irrelevant terms.

\section{Conclusions}

In the context of the conformal field theory, we have considered the energy spectrum of TLL to investigate transport properties in the long-wave and low-temperature limit. We have discussed the role of the irrelevant perturbation $\cos (2\phi)$, representing the Umklapp term, and the role of the flux $\Phi$ insertion, demonstrating that Kubo's linear response theory cannot reproduce the results of the Bethe ansatz and the twisted boson theory. It is shown that this difference should be ascribed to a different role that the Umklapp term plays for the system with or without the flux.

In the absence of the Umklapp term and without the flux insertion, the system is characterized by the two-fold degeneracy related to the momentum inversion $m\rightarrow -m$. With the Umklapp term, this degeneracy is properly treated by considering the symmetric and the antisymmetric combination of states, $|m\rangle \pm |-m\rangle$. Using in the next step Kubo's perturbative formulation to include the effect of the flux, one obtains an additional level splitting that is in contradiction with other approaches.

On the other hand, within the twisted boson theory, because of the nature of flux that behaves as the momentum shift, $\pm m\rightarrow\pm m+\Phi/2\pi$, with the flux there is no two-fold degeneracy. Consequently, the treatment of the Umklapp term within the standard perturbation theory gives the correct description of this irrelevant perturbation. Thus, it is crucial to take into account the irrelevant terms (Umklapp) after the marginal terms (flux) are taken first. Namely, the marginal effects (flux) change the scaling properties of theories \cite{ALCARAZ1988280,Kitazawa_1997}. 

Similar arguments may give a natural reason why the flux effects may be treated perturbatively for gapped systems. In general, these systems may be regarded as a system flowed by relevant perturbations. In this context, the flux as the marginal perturbation loses its relevance, i.e., the energy of the system becomes stable against the flux \cite{Watanabe:2018ahw}. 

Applying these findings, in the context of the twisted boson theory we were able to calculate the Drude weight scaling behaviors for the problem of the XXZ chain, being characterized by the nonanalyticity of the energy for $\Phi=0$. In particular, we find the leading power-law corrections in the low-temperature and the long-wave limit. Our results reveal the difficulties to obtain such behaviors from the opposite side, by using exact calculations for systems of limited sizes and by applying the finite-size analysis. Namely, we find that the low temperature and the long-wave scalings inevitably involve significant finite-size effects.

Finally, it may also be interesting to consider higher-dimensional analogs. If one assumes that a conductor should be described by a massless quantum field theory with irrelevant perturbations, as in the present 1D case a similar divergence of energy under finite flux and for the large system size may exist. Hence, to consider response functions properly, one needs to consider appropriate generalizations of quantum field theories when both, the irrelevant terms and the flux, are present.

\section{Acknowledgement}

YF thanks the previous collaborations related to this project with Ryohei Kobayashi, Masaki Oshikawa, Yuya Nakagawa, and Yuan Yao. YF also thanks helpful comments and discussion with Hosho Katsura, Kiyohide Nomura, Masahiko G. Yamada, and Haruki Watanabe. 

YF acknowledges the support by the QuantiXLie Center of Excellence, a project co-financed by the Croatian Government and European Union through the European Regional Development Fund - the Competitiveness and Cohesion Operational Programme (Grant KK.01.1.1.01.0004). O.S.B. acknowledges the support by Croatian Science Foundation Project No. IP-2016-06-7258.

\appendix


\section{Kubo's response theory and the adiabatic limit}
\label{Kubo_appendix}

We consider Schr\"{o}dinger equation in adiabatic flux insertion. Our discussion in this section follows that of Refs.~\cite{2020arXiv200310390W}.

First, let us introduce the general form of Schr\"{o}dinger equation with Hamiltonian $\hat{H}(\Phi_{t})$,

\begin{equation}
ih\partial_{t}|\alpha(t)\rangle =\hat{H}(\Phi_{t})|\alpha(t)\rangle,
\end{equation}

\noindent where $\Phi_{t}$ is the time dependent flux. In general, the eigenfunction of time-dependent Hamiltonian $H(\Phi_{t})$ cannot give the solution to this equation. To avoid this problem, we concentrate on the adiabatic flux insertion process.

In adiabatic limit, one can obtain the following relations,

\begin{align}
\partial_{t}|k\rangle_{\Phi_{t}}&\sim 0, \\
\hat{H}(\Phi_{t})|k\rangle_{\Phi_{t}} &=E_{k}^{\text{ex}}(\Phi_{t})|k\rangle_{\Phi_{t}}\;,
\end{align}

\noindent where $|k\rangle_{\Phi_{t}}$ is the eigenstate of the Hamiltonian with flux $\Phi_{t}$.
In this limit, one may approximate the solution as eigenfunction of time-dependent Hamiltonian (and we can evaluate it by Bethe ansatz and field theory). Hence, starting from equiribrium mixed state, we can obtain the follwing density matrix,

\begin{equation}
\rho_{t}=\sum_{k}p_{k} |k\rangle_{\Phi_{t}}\langle k|_{\Phi_{t}}, 
\end{equation}

\noindent where $p_{k}$ is the Boltzmann weight.

 It should be noted that we have assumed that energy eigenstate changes continuously under flux insertion in this argument. We start from zero flux $\Phi_{0}= 0$ at $t=0$. Then we sufficiently slowly insert flux to the system until $t=T$ with $\Phi_{T}=\Phi$. Then, we can obtain the response theory as,

\begin{equation}
\begin{split}
&\langle j (T) \rangle-\langle j (0) \rangle=L\partial_{\Phi}\langle H(\Phi_{T})\rangle-L\partial_{\Phi}\langle H(\Phi_{0})\rangle \\
&=\sum_{l\ge 2,k}L\frac{p_{k}}{(l-1)!}\Phi^{l-1}\frac{d^{l}E^{\text{ex}}_{k}}{d\Phi^{l}}(\Phi=0).
\end{split}
\end{equation}

In this traditional argument, we have assumed Taylor's expansion of energy around $\Phi=0$.
Hence, by assuming this Taylor's expansion (or the response theory) and the perturbation theory in flux $\Phi$, we can obtain Eq.~\eqref{naive} without considering the sum rule or other complicated calculations, as shown in \cite{2020arXiv200310390W}.
However, as we have shown in the main text, Taylor's expansion around $\Phi=0$ and Kubo's argument are not consistent with the energy spectrum of the model for finite $\Phi$ obtained by Bethe ansatz. Similar discontinuity can be found in \cite{1998NuPhB.522..533L,1998JPhA...31.9983N}.

Besides this ambiguity of Kubo's argument, we point out some typos of the two quantities,
$\langle k|\partial_{\Phi}^{l}\hat{H}|k\rangle $ and  $\partial_{\Phi}^{l}E_{k}^{\text{ex}}=\partial_{\Phi}^{l}\langle k|\hat{H}(\Phi) |k\rangle$ in Refs.~\cite{2020arXiv200310390W}.
Actually, these quantities are different except for the free fermion point. In other words, the higher derivative extension of the Feynmann-Hellman equation is generally invalid.


\section{Conformal perturbation theory in Hamiltonian formalism}
\label{CFT_appendix}

We review the CFT in Hamiltonian formalism. First, let us assume that the total system is described by the following Hamiltonian,

\begin{equation}
H_{QFT}=H_{CFT}+\frac{\lambda}{2\pi}\int dx \phi_{i}\;,
\end{equation}

\noindent where $H_{CFT}$ is a Hamiltonian of a conformal field theory, $\phi_{i}$ is a primary field of it labeled by index $i$, and $\lambda$ is a coupling constant of the model.
In \cite{CARDY1986186}, the matrix element of the field is calculated as

\begin{equation}
\langle j| \phi_{i} |k \rangle =C_{jik}\left( \frac{2\pi}{L}\right)^{h_{i}}\;,
\end{equation}

\noindent where $C_{ijk}$ is the coefficient of operator product expansion of the theory.

The first- and the second-order perturbation theory gives the following energy spectrum,

\begin{equation}
E_{j}\rightarrow E_{j}+\lambda C_{jij}\left(\frac{2\pi}{L}\right)^{h_{i}-1}+\lambda^{2}\sum_{k}\frac{C_{jik}C_{kij}}{E_{j}-E_{k}}\left( \frac{2\pi}{L}\right)^{2h_{j}-2}
\end{equation}

\noindent where we have used the notation $E_{j}-E_{0}=2\pi h_{i}/L$. One may see that $h_{i}=2$ gives a marginal effect to the energy spectrum by comparing the first and the second term. Hence this formalism may be consistent with the usual Lagrangian formulation of RG analysis.

By applying this formalism to TLL, one can obtain the second-order perturbation theory in \eqref{perturbed_energy}. It should be noted that $\text{cos}2\phi =(e^{2i\phi}+e^{-2i\phi})/2$, and $e^{\pm 2i\phi}$ are the primary fields of the theories. In TLL, the label of a primary field is characterized by the two indexes $m$ and $n$, and the OPE coefficients are obtained by considering $U(1)$ charge of the theory. Hence we can obtain the following of the matrix elements as

\begin{equation}
\begin{split}
&\langle m,n,N,\overline{N}|e^{-2i\phi}|m',n',N',\overline{N'}\rangle \\
&=\delta_{n,n'}\delta_{N,N'} \delta_{\overline{N},\overline{N}'}\delta_{m+2,m'}\left( \frac{2\pi}{L} \right)^{4K(1-\frac{\Phi}{2\pi})}
\end{split}
\end{equation}

\begin{equation}
\begin{split}
&\langle m,n,N,\overline{N}|e^{2i\phi}|m',n',N',\overline{N'}\rangle \\
&=\delta_{n,n'}\delta_{N,N'} \delta_{\overline{N},\overline{N}'}\delta_{m-2,m'}\left( \frac{2\pi}{L} \right)^{4K(1+\frac{\Phi}{2\pi})}.
\end{split}
\end{equation}

\noindent By considering the integration around spacial direction $\int dx$, we can obtain the matrix element of Umklapp term \eqref{matrix}.


\section{Contrbutions from other irrelevant terms}
\label{TT_appendix}

As can be seen in \cite{1998NuPhB.522..533L, 2012IJMPB..2644009S}, in XXZ chain there exist other irrelevant perturbations, $T^{2}$, which are described by the energy-momentum tensor $T\overline{T}$. In principle, one may consider more irrelevant terms, such as $T^{3}$ and $\overline{T}^{3}$, which can contribute to higher response functions. Compared with the Umklapp term, these terms are unique because they do not break the two-fold degeneracy of the original model, and energy eigenstates are eigenstates of these terms as well.

Here, we calculate the contributions of these terms, $T\overline{T}$,
$T^{2}$, $\overline{T}^{2}$, within the conformal perturbation theory of the twisted boson theory. By applying the calculations in Ref.~\cite{1998NuPhB.522..533L}, the contribution of these terms to the energy may be expressed by

\begin{equation}
\begin{split}
&E_{m,0,N,\overline{N}}^{TT}\left(\Phi \right) \\
&\sim \left(\frac{2\pi}{L}\right)^{3} \lambda_{+}H_{1}\left( E\left(m,N,\Phi\right)\right)H_{1}\left( E\left(m,\overline{N},\Phi\right)\right) \\
&+\left(\frac{2\pi}{L}\right)^{3}\lambda_{-}\left( H_{3}\left(E\left(m,N,\Phi\right)\right)
+H_{3}\left(E\left(m,\overline{N},\Phi\right)\right)\right),
\end{split}
\end{equation}

\noindent where we have introduced the following functions,

\begin{align}
H_{1}(E)&=E-\frac{1}{24}, \\
H_{3}(E)&=E^{4}-\frac{E^{2}}{4}, \\
E(x,y,z)&=\frac{1}{2}K\left(x+\frac{z}{2\pi}\right)^{2}+y,
\end{align}

\noindent and $\lambda_{\pm}$ are determined by coupling constants for these terms. Hence, from these terms, one may obtain the following contribution for Drude weight,

\begin{equation}
\begin{split}
&D^{TT}_{s}= L\sum_{m,N,\overline{N}} g(N)g(\overline{N}) p_{m,0,N,\overline{N}} \frac{d^{2}E_{m,0,N,\overline{N}}^{TT}}{d\Phi^{2}} \\
&\sim \frac{4\pi \left(\lambda_{+}+2\lambda_{-}\right)}{L^{2}}\sum_{m,N,\overline{N}}g(N)g(\overline{N}) \\
&\times\left(3K^{2}m^{2}+K\left(N+\overline{N}-\frac{1}{12}\right) \right) p_{m,0,N,\overline{N}} \\
&+\frac{\pi K}{6L^{2}}\left(\lambda_{+}-2\lambda_{-}\right) \\
&=F_{TT}+Q_{TT}+\frac{\pi K}{6L^{2}}\left(\lambda_{+}-2\lambda_{-}\right),
\end{split}
\end{equation}

\noindent where we have introduced the functions 

\begin{align}
F_{TT}&=\frac{4\pi b}{L^{2}}\sum_{m}m^{2} \frac{e^{-\frac{2\pi \beta K}{L}m^{2}}}{z(L,\beta)}, \\
Q_{TT}&=-\frac{2 b'}{L}\frac{\partial}{\partial \beta}\text{log}\frac{1}{\eta(\tau)^{2}}=\frac{4i b'}{L^{2}}\frac{\partial}{\partial \tau}\text{log}\eta(\tau), 
\end{align}

\noindent and $b$ and $b'$ are parameters defined as $b=3K^{2}\left(\lambda_{+}+2\lambda_{-}\right)$, $b'=2K\left(\lambda_{+}+2\lambda_{-}\right)$. We use the standard notation for the Riemann theta function, $z(L,\beta)=\theta_{0,0}( i\beta K/L,0)$.

By using the following relation for the Riemann theta function and the logarithmic degivative,

\begin{align}
\partial_{\tau} \theta_{00} (t ,z)&= \frac{i}{4\pi}\partial_{z}^{2} \theta_{00} (t ,z),\\
\partial_{z}\text{log} \theta_{00}(t,z)&=4\pi \sum_{k=1} ^{\infty}\left(-1\right)^{k}\frac{e^{k\pi i t}}{1-e^{2k\pi i t}}\text{sin}2\pi k z
\end{align}

\begin{equation}
\frac{f''}{f}=\left(\frac{f'}{f}\right)'+\left(\frac{f'}{f}\right)^{2} 
\end{equation}

\noindent one obtains

\begin{equation}
F_{TT}\sim \frac{4\pi b}{L^{2}} \sum_{k=1}^{\infty} \left(-1\right)^{k}\frac{ke^{k\pi i t}}{1-e^{2k\pi i t}}
\end{equation}

\noindent where $t=2i\beta K/L$.

Hence for low temperatures, we may use the approximation $1-e^{2k\pi i t}\sim 1$, and $F_{TT}$ may be expressed by

\begin{equation}
F_{TT}\sim\frac{4\pi b}{L^{2}} \frac{e^{\pi it}}{\left(1+e^{\pi i t}\right)^{2}}\sim \frac{1}{L^{2}}e^{-\frac{2\pi\beta K}{L}}
\end{equation}

For finite temperatures, one may use an approximation to replace $\sum$ and $\int$. We obtain

\begin{equation}
F_{TT} \sim\frac{4\pi b}{L^{2}}\frac{\int dx x^{2}e^{-\frac{2\pi\beta K}{L}x^{2}}}{\int dx e^{-\frac{2\pi\beta K}{L}x^{2}}}\sim\frac{2 b}{L\beta K}\;.
\end{equation}

\appendix


\bibliography{Ref_drude}

\end{document}